# Quantum Field and Cosmic Field

## Finite Geometrical Field Theory of Matter Motion Part Three

Xiao Jianhua

Natural Science Foundation Research Group, Shanghai Jiaotong University

Shanghai, P.R.C

**Abstract:** Quantum field is well established and confirmed by experiments. This research establishes an operational measurement way to express the quantum field theory in a geometrical form. Based on the relationship among the space gauge and time gauge field, a finite geometrical field motion equation of quantum can be established based on the least action principle. This may be valuable for understanding the underlying concepts of quantum matter motion. In four-dimensional spacetime continuum, the orthogonal rotation is defined. It forms two sets of equations: one set is geometrical equations, another set is the motion equations. The Lorentz transformation can be directly derived from the geometrical equations, and the proper time of general relativity is well expressed by time displacement field. By the motion equations, the typical time displacement field of matter motion is discussed. The research shows that the quantum field theory can be established based on the concept of orthogonal rotation. On this sense, the quantum matter motion in physics is viewed as the orthogonal rotation (a special deformation) of spacetime continuum caused by basic solution of time displacement. In this paper, it shows that there are three typical quantum solutions. One is particle-like solution, one is generation-type solution, and one is pure wave type solution. For each typical solution, the force fields are different. When the quantum solutions are given, the force fields may have different features. Many features of quantum field can be well explained by this theoretic form. Finally, the quantum field, gravity field, and electrical field are considered in an united form, temporally the united field is named as cosmic field as in essence it will leads to the concept of absolute spacetime. The general matter motion is discussed, the main conclusions are: (1). Geometrically, cosmic vacuum field can be described by the curvature spacetime, such as Minkowski world; (2). The spatial deformation of planet is related with a planet electromagnetic field. Further more, the research shows that the volume variation will cause the harmonic vibration of time displacement; (3). For electric charge less matter, the volume of matter will be expanding infinitely. This may be the intrinsic explanation about the Big-ban when the matter is in cosmic background (scale); (4).For strong electric charge matter, it shows that the volume of matter will be contracting infinitely. This may be the intrinsic explanation about the Black-hole when the matter is in cosmic background (scale).

**Key Words:** quantum mechanics, general relativity, motion equation, orthogonal rotation

## 1. Introduction

In the papers "Inertial System and Special Relativity" [1] and "Gravity Field and Electromagnetic Field" [2], the symmetric matter motion transformation and pure time displacement field are discussed. In this paper, the anti-symmetric motion transformation will be discussed.

For any rank-two tensor, it can be divided into the sum of a symmetric tensor and an anti-symmetric tensor. According to our previous research results, the symmetric tensor corresponds to Newtonian matter motion. The pure time displacement field is a non-symmetric tensor, it has been treated as the conservative and electromagnetic fields. For the matter motion in the general motion forms, as the symmetric part corresponds to Newtonian matter motion discussed earlier, this part will discusses the anti-symmetric form of matter motion tensor.

Based on Chen Zhida's research, the anti-symmetric tensor corresponds to an orthogonal rotation tensor [3-4]. According to this formulation mechanism, a general rank-two tensor $F_j^i$ in three dimensional space, say the motion transformation tensor, can be expressed by the gradient of displacement $u^i$ as following:

$$F_j^i = u^i\big|_j + \delta_j^i \tag{2}$$

where, $u^i\big|_j$ express the covariant derivation of displacement fields; $\delta_j^i$ is Kronecker-delta. Note that for three dimensional space $i = 1,2,3$.

Chen Z D has shown that the transformation can be decomposed into the addition of one symmetry tensor expressing stretching and one unit orthogonal tensor expressing local rotation [3-4]. That is:

$$F_j^i = S_j^i + R_j^i \tag{3}$$

Where:

$$S_j^i = \frac{1}{2}(u^i\big|_j + u^i\big|_j^T) - (1-\cos\Theta)L_k^i L_j^k \tag{4}$$

$$R_j^i = \delta_j^i + \sin\Theta \cdot L_j^i + (1-\cos\Theta)L_k^i L_j^k \tag{5}$$

$$L_j^i = \frac{1}{2\sin\Theta}(u^i\big|_j - u^i\big|_j^T) \tag{6}$$

$$\sin\Theta = \frac{1}{2}[(u^1\big|_2 - u^2\big|_1)^2 + (u^2\big|_3 - u^3\big|_2)^2 + (u^3\big|_1 - u^1\big|_3)^2]^{\frac{1}{2}} \tag{7}$$

The parameter $\Theta$ represents local average rotation angular, $L_j^i$ represents the local average rotation direction tensor, $R_j^i$ is an unit-orthogonal rotation tensor.

For large rotation, this decomposition can be extended to the form [5]:

$$F_j^i = \tilde{S}_j^i + \frac{1}{\cos\theta}\tilde{R}_j^i \tag{8}$$

where,

$$\tilde{S}_j^i = \frac{1}{2}(u^i\big|_j + u^i\big|_j^T) - (\frac{1}{\cos\theta}-1)(\tilde{L}_k^i \tilde{L}_j^k + \delta_j^i) \tag{9}$$

$$(\cos\theta)^{-1}\tilde{R}_j^i = \delta_j^i + \frac{\sin\theta}{\cos\theta}\tilde{L}_j^i + (\frac{1}{\cos\theta}-1)(\tilde{L}_k^i \tilde{L}_j^k + \delta_j^i) \tag{10}$$

$$\tilde{R}_j^i = \delta_j^i + \sin\theta \cdot \tilde{L}_j^i + (1-\cos\theta)\tilde{L}_k^i \tilde{L}_j^k \tag{11}$$

$$\tilde{L}_j^i = \frac{\cos\theta}{2\sin\theta}(u^i\big|_j - u^i\big|_j^T) \tag{12}$$

$$(\cos\theta)^{-2} = 1 + \frac{1}{4}[(u^1\big|_2 - u^2\big|_1)^2 + (u^2\big|_3 - u^3\big|_2)^2 + (u^3\big|_1 - u^1\big|_3)^2] \tag{13}$$

The parameter $\theta$ represents local average rotation angular, $\tilde{L}_j^i$ represents the local average rotation direction tensor, $\tilde{R}_j^i$ is an unit-orthogonal rotation tensor.

In this paper, the concept of orthogonal rotation will be extended to four-dimensional spacetime continuum for simple cases in an operational way. To get its intrinsic meaning please refer the Appendix A (*Mathematic Feature of Deformation Tensor*) for three-dimensional case.

This idea of orthogonal rotation in four-dimensional spacetime is well supported by Lorentz transformation, as the Lorentz transformation forms a rotational invariant group in four-dimensional spacetime on mathematic sense. However, as the previous research has shown, the Lorentz transformation is correct conditionally and cannot be the intrinsic feature to express such a kind of general rotation of matter motion.

For our purpose, the Lorentz transformation will be derived from the least action principle for orthogonal rotation motion transformation. Firstly, the basic equations are reviewed simply.

The matter motion in four-dimensional spacetime continuum is defined by:

$$\begin{vmatrix} \vec{g}_1 \\ \vec{g}_2 \\ \vec{g}_3 \\ \vec{g}_4 \end{vmatrix} = \begin{vmatrix} 1+u^1\big|_1 & u^2\big|_1 & u^3\big|_1 & u^4\big|_1 \\ u^1\big|_2 & 1+u^2\big|_2 & u^3\big|_2 & u^4\big|_2 \\ u^1\big|_3 & u^2\big|_3 & 1+u^3\big|_3 & u^4\big|_3 \\ u^1\big|_4 & u^2\big|_4 & u^3\big|_4 & 1+u^4\big|_4 \end{vmatrix} \cdot \begin{vmatrix} \vec{g}_1^0 \\ \vec{g}_2^0 \\ \vec{g}_3^0 \\ \vec{g}_4^0 \end{vmatrix} \tag{14}$$

That is the initial basic vectors and current basic vectors meet transformation equation:

$$\vec{g}_i = F_i^{\ j}\vec{g}_j^0 \tag{15}$$

The matter motion transformation tensor $F_j^i$ is determined by equation:

$$F_i^{\ j} = u^j\big|_i + \delta_i^j \tag{16}$$

where, $\big|_i$ represents covariant derivative with respect to coordinator $x^i$, $\delta_i^j$ is an unit tensor.

The displacement field $u^i$ is also defined in the initial four-dimensional co-moving coordinator system. Therefore, the matter motion is expressed by displacement field $u^i$ measured in standard physical measuring system. Its three dimenssional case has been studied by Chen Zhida [3-4].

Generally, the matter motion transformation tensor is non-commutative. Its covariant index and contra-variant index are defined respects with to the initial four-dimensional co-moving coordinator system.

The action for matter motion is defined as:

$$Action = \int W(F_j^i) dx^1 dx^2 dx^3 dx^4 \tag{17}$$

where, $W$ is a general function.

The least action principle [6-7] gives the field motion equations for contra-covariant force as:

$$\sigma^i_j\big|_j = f^i \tag{18}$$

where,

$$\sigma^i_j = C^{ik}_{jl}(F^l_k - \tilde{F}^l_k) \tag{19}$$

$$C^{ik}_{jl} = \frac{\partial^2 W}{\partial(\tilde{F}^i_j)\partial(\tilde{F}^k_l)} \tag{20}$$

$$\frac{\partial W}{\partial(\tilde{F}^i_j)}\bigg|_j = -f^i \tag{21}$$

However, if one does care how the original matter motion state is in the absolute spacetime, then the initial matter itself can be generally defined as the initial co-moving coordinator system. In this case, $F^i_j - \tilde{F}^i_j$ defines the incremental motion transformation. Of course, the selection of the initial co-moving coordinator system can make if the reference matter is isotropic:

$$\tilde{F}^i_j = \delta^i_j \tag{22}$$

In this selection, $Action(0) = \int W(\delta^i_j) dx^1 dx^2 dx^3 dx^4$ defines the action of reference matter respect with vacuum or ether.

On this sense, the initial matter is taken to define the spacetime continuum where the matter incremental deformation under discussion is taking place. Of course, by such a selection, the gauge $g^0_{ij}$ is determined by our selection. Therefore, $g^0_{ij}$ is arbitrary and has eight independent parameters which are completely determined by the reference matter state.

The principle of physical laws covariant invariance requires the general reference matter motion be coordinator independent. To meet these requirements, the tensor $C^{ik}_{jl}$ must be isotropic tensor. Therefore, one has:

$$C^{ik}_{jl} = \lambda \delta^i_j \delta^k_l + \mu \delta^i_l \delta^k_j \tag{23}$$

where, $\lambda$ and $\mu$ are the intrinsic feature of matter referring to reference matter. For Newton mechanics, the reference matter is vacuum or ether. For the isotropic tensor in mixture form please refer [8]. In standard physical theory, the mixed form is refused. However, the research shows that the mixture form is necessary to show the non-commutative feature of matter motion transformation tensor. (refer, Appendix A (*Mathematic Feature of Deformation Tensor*)).

When the tensor $F^i_j - \tilde{F}^i_j$ has non-symmetric components, no matter what coordinator system is selected as the initial co-moving coordinator system, the tensor $F^i_j - \tilde{F}^i_j$, in general case, has eight independent intrinsic components. In this case, the motion equation (18) is not enough to determine the tensor $F^i_j - \tilde{F}^i_j$. For three-dimensional space, Chen Zhida's research [3-4] shows that the another motion equation is:

$$\sigma^i_j\big|_i = f_j \tag{24}$$

For four dimensional spacetime, the equation should be true also. Note that here the covariant force is defined as:

$$\frac{\partial W}{\partial (\tilde{F}^i_j)}\bigg|_i = -f_j \tag{25}$$

This equation defines the covariant force. On this sense, the motion equation (24) is the contra-covariant form motion equation.

Based on Chen Zhida's research, geometrically, the equation (18) corresponds to the total force along a special direction while the equation (24) corresponds to the total force on a special surface. For deformable matter space, they are not the same. Only for the standard Cartesian space, they are the same.

For a given matter, to make the equation (22) be effective, the initial co-moving coordinator system must be a general Riemannian spacetime. In this case, according to the definition equation (15) there are relation equations:

$$f_j = g^0_{ij} f^i \tag{26}$$

Summering above researches, the motion equations for non-symmetric motion transformation are:

$$\sigma^i_j\big|_j = f^i \tag{27-1}$$

$$\sigma^i_j\big|_i = f_j \tag{27-2}$$

$$\sigma^i_j = C^{ik}_{jl}(F^l_k - \tilde{F}^l_k) \tag{27-3}$$

where, the matter feature and the action forces referring to the reference matter (defined by motion transformation $\tilde{F}^i_j$) are given by:

$$\frac{\partial W}{\partial (\tilde{F}^i_j)}\bigg|_j = -f^i \tag{28-1}$$

$$\frac{\partial W}{\partial (\tilde{F}^i_j)}\bigg|_i = -f_j \tag{28-2}$$

$$C^{ik}_{jl} = \frac{\partial^2 W}{\partial (\tilde{F}^i_j)\partial (\tilde{F}^k_l)} \tag{28-3}$$

When the reference matter is vacuum or ether, that is if $\tilde{F}^i_j = \delta^i_j$, the isotropic feature of cosmic spacetime continuum will has:

$$C^{ik}_{jl} = \lambda \delta^i_j \delta^k_l + \mu \delta^i_l \delta^k_j \tag{29}$$

Based on the motion equations, the paper will firstly discuss the orthogonal rotation of rigid matter bady in four-dimensional spacetime. The results show that the Lorentz transformation is the natural results if motion is measured in inertial motion concept. Further more, the research shows that the Newtonian moment space is equivalent with the wave-number space, so this basic relations between the classical mechanics and the quantum mechanics is naturally derived from the least action principle for orthogonal rotation motion. Secondly, the orthogonal rotation for space-deformable matter is discussed, which leads to the non-linear equation of quantum mechanics. Finally, the basic four force in nature are discussed in details to show the basic features of cosmic field, which is defined by the general form of motion transformation.

**2. Matter Rotation in Four-dimensional Spacetime**

For incremental deformation $F_j^i = S_j^i + R_j^i$, when the symmetric tensor is zero the incremental deformation becomes an unit orthogonal rotation tensor, the current gauge field $g_{ij}$ is the same as the initial gauge field defined by the reference matter. That is:

$$g_{ij} = F_i^k F_j^k = R_i^k R_j^k = \delta_{ij} \tag{30}$$

For matter rotation in four-dimensional spacetime continuum, the motion transformation is:

$$F_j^i = \begin{vmatrix} R_1^1 & R_1^2 & R_1^3 & u^4|_1 \\ R_2^1 & R_2^2 & R_2^3 & u^4|_2 \\ R_3^1 & R_3^2 & R_3^3 & u^4|_3 \\ u^1|_4 & u^2|_4 & u^3|_4 & 1+u^4|_4 \end{vmatrix} \tag{31}$$

Where the $R_j^i$ space components for are defined by the (3)-(7) equations. Hence, for isotropic reference matter, according to this equation, the simplest unit orthogonal rotation introduced by equation (30) is defined by the following conditions:

$$R_4^k R_j^k = u^1|_4 \cdot R_j^1 + u^2|_4 \cdot R_j^2 + u^3|_4 \cdot R_j^3 + (1+u^4|_4) \cdot u^4|_j = 0, \text{ for } j=1,2,3 \tag{32}$$

$$R_4^k R_4^k = u^1|_4 \cdot u^1|_4 + u^2|_4 \cdot u^2|_4 + u^3|_4 \cdot u^3|_4 + (1+u^4|_4) \cdot (1+u^4|_4) = 1 \tag{33}$$

For the idea reference matter such as vacuum, the $dx^4 = cdt$ can be introduced to get the physical components. They can be rewritten as by physical components as:

$$V^1 \cdot R_j^1 + V^2 \cdot R_j^2 + V^3 \cdot R_j^3 = -c^2 \cdot (1+\frac{\partial \tilde{u}^4}{\partial t}) \cdot \frac{\partial \tilde{u}^4}{\partial x^j}, \text{ for } j=1,2,3 \tag{34}$$

$$(V^1)^2 + (V^2)^2 + (V^3)^2 = c^2 \cdot [1-(1+\frac{\partial \tilde{u}^4}{\partial t})^2] \tag{35}$$

The striking feature is that once the spatial velocity $V^i$ and rotation tensor is measured the time displacement is completely determined.

On the other hand, by the definition of orthogonal rotation, one also has:

$$R_i^4 R_i^j = u^4|_1 \cdot R_1^j + u^4|_2 \cdot R_2^j + u^4|_3 \cdot R_3^j + (1+u^4|_4) \cdot u^j|_4 = 0, \text{ for } j=1,2,3 \tag{36}$$

$$R_i^4 R_i^4 = u^4|_1 \cdot u^4|_1 + u^4|_2 \cdot u^4|_2 + u^4|_3 \cdot u^4|_3 + (1+u^4|_4) \cdot (1+u^4|_4) = 1 \tag{37}$$

Their physical components forms are:

$$\frac{\partial \tilde{u}^4}{\partial x^1} \cdot R_1^j + \frac{\partial \tilde{u}^4}{\partial x^2} \cdot R_2^j + \frac{\partial \tilde{u}^4}{\partial x^3} \cdot R_3^j = -\frac{1}{c^2}(1+\frac{\partial \tilde{u}^4}{\partial t}) \cdot V^j, \text{ for } j=1,2,3 \quad (38)$$

$$(\frac{\partial \tilde{u}^4}{\partial x^1})^2 + (\frac{\partial \tilde{u}^4}{\partial x^2})^2 + (\frac{\partial \tilde{u}^4}{\partial x^3})^2 = \frac{1}{c^2}[1-(1+\frac{\partial \tilde{u}^4}{\partial t})^2] \quad (39)$$

Here, the $\tilde{u}^4 = cu^4$ is the physical component of time displacement. The striking feature is that once the spatial gradient of time displacement $\frac{\partial \tilde{u}^4}{\partial x^i}$ ($i=1,2,3$) and spatial rotation tensor is measured the spatial velocity is completely determined.

From spatial consideration, once the spatial velocity is given, the spatial orthogonal rotation is determined (only need three independent parameters). Therefore, the matter rotation in four dimensional spacetime continuum can be described in velocity space $V^i$ or in the spatial gradient of time displacement $\frac{\partial \tilde{u}^4}{\partial x^i}$ ($i=1,2,3$) space (which is named as wave-number space). So, at the following discussion such a kind of matter motion will be named as quantum motion.

Comparing with the practice in classical quantum mechanics, the former is called space terminology, the later is called wave terminology [9].

Define:

$$v_i = \frac{\partial \tilde{u}^4}{\partial x^i}, (i=1,2,3) \quad (40)$$

In Cartesian initial coordinator system, the related equations can be simplified as:

$$V^l \cdot R_j^l = -c^2 \cdot (1+\frac{\partial \tilde{u}^4}{\partial t}) \cdot v_j, \text{ for } j,l=1,2,3 \quad (41)$$

$$V^2 = c^2 \cdot [1-(1+\frac{\partial \tilde{u}^4}{\partial t})^2] \quad (42)$$

$$v_l R_l^j = -\frac{1}{c^2}(1+\frac{\partial \tilde{u}^4}{\partial t}) \cdot V^j, \text{ for } j,l=1,2,3 \quad (43)$$

$$v^2 = \frac{1}{c^2}[1-(1+\frac{\partial \tilde{u}^4}{\partial t})^2] \quad (44)$$

They form the intrinsic geometrical equations for quantum mechanics in finite geometrical field theory of matter motion.

## 3. Motion Equation of Quantum Mechanics

For initial isotropic spacetime continuum, the intrinsic feature of matter is defined by isotropic tensor $C_{jl}^{ik}$. That is: $C_{jl}^{ik} = \lambda \delta_j^i \delta_l^k + \mu \delta_l^i \delta_j^k$. The stress field is:

$$\sigma_j^i = \lambda(F_l^l - \delta_l^l)\delta_j^i + \mu(F_j^i - \delta_j^i) \quad (45)$$

For pure matter rotation in four-dimensional spacetime, as: $S_j^i = 0$, one has:

$$F_j^i = S_j^i + R_j^i = R_j^i \quad (46)$$

So, the stress field in physical components form is:

$$\sigma_j^i = \lambda(R_1^1 + R_2^2 + R_3^3 + \frac{\partial \tilde{u}^4}{\partial t} - 3)\delta_j^i + \mu \begin{vmatrix} R_1^1-1 & R_1^2 & R_1^3 & cv_1 \\ R_2^1 & R_2^2-1 & R_2^3 & cv_2 \\ R_3^1 & R_3^2 & R_3^3-1 & cv_3 \\ V^1/c & V^2/c & V^3/c & \frac{\partial \tilde{u}^4}{\partial t} \end{vmatrix} \quad (47)$$

The motion equations are:

$$\lambda \frac{\partial R_j^j}{\partial x^i} + \lambda \frac{\partial^2 \tilde{u}^4}{\partial t \partial x^i} + \mu \frac{\partial R_j^i}{\partial x^j} + \frac{\mu}{c^2} \frac{\partial V^i}{\partial t} = f^i, \quad i,j=1,2,3 \quad (48\text{-}1)$$

$$\frac{\lambda}{c} \frac{\partial R_j^j}{\partial t} + \mu c \frac{\partial v_j}{\partial x^j} + \frac{\lambda+\mu}{c} \frac{\partial^2 \tilde{u}^4}{(\partial t)^2} = f^4, \quad j=1,2,3 \quad (48\text{-}2)$$

And:

$$\lambda \frac{\partial R_j^j}{\partial x^i} + \lambda \frac{\partial^2 \tilde{u}^4}{\partial t \partial x^i} + \mu \frac{\partial R_i^j}{\partial x^j} + \mu \frac{\partial v_i}{\partial t} = f_i, \quad i,j=1,2,3 \quad (49\text{-}1)$$

$$\frac{\lambda}{c} \frac{\partial R_j^j}{\partial t} + \frac{\mu}{c} \frac{\partial V^j}{\partial x^j} + \frac{\lambda+\mu}{c} \frac{\partial^2 \tilde{u}^4}{(\partial t)^2} = f_4, \quad j=1,2,3 \quad (49\text{-}2)$$

Please note that the equation (48-2) is similar with the Schrodinger equation in essential sense. To clear their meaning, three simple typical cases are discussed below.

*(A). Newtonian Matter Rotation*

If time displacement field related items are higher order infinitesimals, the motion equations for Newtonian matter in three dimensional space are (note that $\rho = \mu/c^2$ defines the Newtonian mass [1]):

$$\lambda \frac{\partial R_j^j}{\partial x^i} + \mu \frac{\partial R_j^i}{\partial x^j} = -\rho \frac{\partial V^i}{\partial t} + f^i, \quad i,j=1,2,3 \quad (50\text{-}1)$$

$$\frac{\lambda}{c} \frac{\partial R_j^j}{\partial t} = f^4, \quad j=1,2,3 \quad (50\text{-}2)$$

$$\lambda \frac{\partial R_j^j}{\partial x^i} + \mu \frac{\partial R_i^j}{\partial x^j} = f_i, \quad i,j=1,2,3 \quad (50\text{-}3)$$

$$\frac{\lambda}{c} \frac{\partial R_j^j}{\partial t} + \frac{\mu}{c} \frac{\partial V^j}{\partial x^j} = f_4, \quad j=1,2,3 \quad (50\text{-}4)$$

The equation (50-1) is the Chen's form of Euler equations for local average rotation of pure rotational deformation of continuum in three dimensional space [3-4]. As $R_l^l = 1 + 2\cos\Theta$, the equation (50-2) shows that the rotation angular time variation is determined by cosmic force. This phenomena can be observed or be used to explain the planet rotation with the rotation angular time variation.

By equation (50-1) and (50-2), one gets:

$$\mu(\frac{\partial R_i^j}{\partial x^j} - \frac{\partial R_j^j}{\partial x^j}) = \rho \frac{\partial V^i}{\partial t} + f_i - f^i \quad (51\text{-}1)$$

This is the shear wave equation for elastic continuum, however please note that for classical elastic continuum the parameter $(\lambda,\mu)$ are defined on the matter itself at initial state. As here, the parameters are defined respect with vacuum or ether, the equation (51-1) describes the planet shear motion, that is, the variation of planet rotation in cosmic background.

The equation (50-4) shows that the relation between the time variation of rotation angular and the volume variation is determined by cosmic force. It relates the spatial volume variation with the rotation angular variation. This can be used to explain that how the planet volume variation is companied with rotation angular velocity.

By equation (50-2) and (50-4), one gets:

$$\mu \frac{\partial V^j}{\partial x^j} = f_4 - f^4 \tag{51-2}$$

Recalling that $\mu$ is related with Newtonian mass quantity, the $\lambda + \mu$ is related with Coulomb charge, it can be inferred that the Newtonian matter rotation not only depends on its Newtonian mass but also depends on its Coulomb charge.

When $f^i = f_i$, $i = 1,2,3,4$, the equation (51) becomes Euler rigid rotation equation.

Summering above results about Newtonian matter rotation in vacuum or ether, it can be concluded that the Euler rotation equation for rigid body is the only rotation mode when the reference matter is vacuum or Cartesian spacetime. Therefore, for curvature spacetime, that is, when the real cosmic field is in consideration, the equations (50) can be used. Therefore, this results may be valuable for astronomy research and cosmics study. In fact, the equation (50) can be used to explain the so-called self rotation of spaceship, and so can be used to control to flight of spaceship.

*(B). Conservative Matter Rotation*

If the space Newtonian acceleration is higher infinitesimal and the matter has no volume variation, the motion equation becomes:

$$\lambda \frac{\partial R_j^j}{\partial x^i} + \lambda \frac{\partial^2 \tilde{u}^4}{\partial t \partial x^i} + \mu \frac{\partial R_j^i}{\partial x^j} = f^i, \quad i,j = 1,2,3 \tag{52-1}$$

$$\frac{\lambda}{c} \frac{\partial R_j^j}{\partial t} + \mu c \frac{\partial v_j}{\partial x^j} + \frac{\lambda + \mu}{c} \frac{\partial^2 \tilde{u}^4}{(\partial t)^2} = f^4, \quad j = 1,2,3 \tag{52-2}$$

And:

$$\lambda \frac{\partial R_j^j}{\partial x^i} + \lambda \frac{\partial^2 \tilde{u}^4}{\partial t \partial x^i} + \mu \frac{\partial R_i^j}{\partial x^j} + \mu \frac{\partial v_i}{\partial t} = f_i, \quad i,j = 1,2,3 \tag{52-3}$$

$$\frac{\lambda}{c} \frac{\partial R_j^j}{\partial t} + \frac{\lambda + \mu}{c} \frac{\partial^2 \tilde{u}^4}{(\partial t)^2} = f_4, \quad j = 1,2,3 \tag{52-4}$$

Recalling the related definition in "Gravity Field and Electromagnetic Field" [2], when the time gradient is commutative the matter is in conservative form. So, for conservative field, the equation can be rewritten as (note equation (40)):

$$\lambda \frac{\partial R_j^j}{\partial x^i} + \mu \frac{\partial R_j^i}{\partial x^j} = -\lambda \frac{\partial^2 \tilde{u}^4}{\partial t \partial x^i} + f^i, \quad i,j = 1,2,3 \tag{53-1}$$

$$\frac{\lambda}{c} \frac{\partial R_j^j}{\partial t} = -\mu c \frac{\partial^2 \tilde{u}^4}{\partial x^j \partial x^j} - \frac{\lambda + \mu}{c} \frac{\partial^2 \tilde{u}^4}{(\partial t)^2} + f^4, \quad j = 1,2,3 \tag{53-2}$$

And:

$$\lambda \frac{\partial R_j^j}{\partial x^i} + \mu \frac{\partial R_i^j}{\partial x^j} = -(\lambda + \mu)\frac{\partial^2 \tilde{u}^4}{\partial t \partial x^i} + f_i, \quad i, j = 1,2,3 \tag{53-3}$$

$$\frac{\lambda}{c} \frac{\partial R_j^j}{\partial t} = -\frac{\lambda + \mu}{c} \frac{\partial^2 \tilde{u}^4}{(\partial t)^2} + f_4, \quad j = 1,2,3 \tag{53-4}$$

By equations (53-1) and (53-3), one gets:

$$\mu(\frac{\partial R_j^i}{\partial x^j} - \frac{\partial R_i^j}{\partial x^j}) = \mu \frac{\partial^2 \tilde{u}^4}{\partial t \partial x^i} + f^i - f_i, \quad i, j = 1,2,3 \tag{54-1}$$

By equations (53-2) and (53-4), one gets:

$$\mu c \frac{\partial^2 \tilde{u}^4}{\partial x^j \partial x^j} = f^4 - f_4 \tag{54-2}$$

When all forces are zero, the conservative matter has a shear-wave-like local rotation, it means that the gravity field (defined by $\frac{\partial u^4}{\partial t} = \frac{M}{r}$), the static electrical field (defined by $\frac{\partial u^4}{\partial t} = \frac{Q}{r}$), and the electric field (defined by $\frac{\partial^2 u^4}{\partial t \partial x^i}$) can be in shear wave form.

### (C). Electromagnetic Field Rotation

If the space Newtonian acceleration is higher infinitesimal and the matter has no volume variation, for non-commutative field, the motion equation becomes:

$$\lambda \frac{\partial R_j^j}{\partial x^i} + \lambda \frac{\partial^2 \tilde{u}^4}{\partial t \partial x^i} + \mu \frac{\partial R_j^i}{\partial x^j} = f^i, \quad i, j = 1,2,3 \tag{55-1}$$

$$\frac{\lambda}{c} \frac{\partial R_j^j}{\partial t} + \mu c \frac{\partial^2 \tilde{u}^4}{\partial x^j \partial x^j} + \frac{\lambda + \mu}{c} \frac{\partial^2 \tilde{u}^4}{(\partial t)^2} = f^4, \quad j = 1,2,3 \tag{55-2}$$

And:

$$\lambda \frac{\partial R_j^j}{\partial x^i} + \lambda \frac{\partial^2 \tilde{u}^4}{\partial t \partial x^i} + \mu \frac{\partial R_i^j}{\partial x^j} + \mu \frac{\partial^2 \tilde{u}^4}{\partial x^i \partial t} = f_i, \quad i, j = 1,2,3 \tag{55-3}$$

$$\frac{\lambda}{c} \frac{\partial R_j^j}{\partial t} + \frac{\lambda + \mu}{c} \frac{\partial^2 \tilde{u}^4}{(\partial t)^2} = f_4, \quad j = 1,2,3 \tag{55-4}$$

By equation (55-1) and (55-3), one gets:

$$\mu(\frac{\partial R_j^i}{\partial x^j} - \frac{\partial R_i^j}{\partial x^j}) = \mu \frac{\partial^2 \tilde{u}^4}{\partial x^i \partial t} + f^i - f_i, \quad i, j = 1,2,3 \tag{56-1}$$

By equation (55-2) and (55-4), one gets:

$$\mu c \frac{\partial^2 \tilde{u}^4}{\partial x^j \partial x^j} = f^4 - f_4 \tag{56-2}$$

When all forces are zero, the electromagnetic field matter has a shear-wave-like local rotation, it means that the magnetic field vector potential (defined by $\frac{\partial u^4}{\partial x^i}$) time variation can be in shear wave form.

### 4. Quantum Field Geometrical Forms

For general case, the quantum matter motion must meet the geometrical equations and motion equations at the same time.

For matter motion in conservative matter field, defined by: $\frac{\partial u^4}{\partial x^4} = \frac{M}{r}$ refereeing to charge source, the matter behaves as quantum field. When the conservative matter is taken as the reference matter, such as the quantum in Earth gravity field, let:

$$\eta = (1 + \frac{\partial \tilde{u}^4}{\partial t}) \quad (57)$$

be a constant. The geometrical equations (40-44) become:

$$V^l \cdot R^l_j = -c^2 \cdot \eta \cdot v_j, \text{ for } j,l = 1,2,3 \quad (57\text{-}1)$$

$$V^2 = c^2 \cdot [1 - \eta^2] \quad (57\text{-}2)$$

$$v_l R^j_l = -\frac{\eta}{c^2} \cdot V^j, \text{ for } j,l = 1,2,3 \quad (57\text{-}3)$$

$$v^2 = \frac{1}{c^2}[1 - \eta^2] \quad (57\text{-}4)$$

By equation (57-2), in conservative matter field, the velocity square is constant. From this point to see, the kinetic energy conservation is only effective for conservative field matter.

From spatial measurement consideration, the equation (57-2) gives:

$$(dx^1)^2 + (dx^2)^2 + (dx^3)^2 - (cdt)^2 = -\eta^2 c^2 (dt)^2 \quad (58\text{-}1)$$

Therefore, the reference matter invariance, that is the invariant of $\eta$, is equivalent with the proper time invariant in general relativity. In fact, this equation explains the true physical meaning of Lorentz transformation. Please remember that the Lorentz electromagnetic wave equation is taken the cosmic background as the reference matter.

From time measurement consideration, the equation (57-4) gives:

$$(\frac{\partial \tilde{u}^4}{\partial x^1})^2 + (\frac{\partial \tilde{u}^4}{\partial x^2})^2 + (\frac{\partial \tilde{u}^4}{\partial x^3})^2 = \frac{1}{c^2}(1 - \eta^2) \quad (58\text{-}2)$$

For general cases, the quantum matter $u^4$ has three typical solutions [1-2].

*(A). Constant Time Speed Matter (Conservative Matter Field Approximation)*

When the time speed is constant, the parameter $\eta$ is constant about time. The equation (57-2) shows that the matter must have an absolute velocity respect with reference matter (vacuum or ether). The equation (57-4) shows that the magnetic vector potential is non zero. In fact, the equation (57-1) shows that the magnetic vector potential is produced by the local rotation of matter. This can be used to explain the magnetic related with the self-rotation of quantum matter.

Taking Newtonian approximation, its motion equation is described by equations (50). Once the macro velocity and rotation are gotten, the equations (57) show that there must exist an static magnetic field. Therefore, any planet must have static magnetic field. Further more, the static magnetic field can exist for particles also.

For conservative field matter, $\eta = 1 + \frac{M}{r}$, the equations (57) can be approximated as:

$$\frac{V^l \cdot R^l_j}{c^2} + v_j = \frac{M}{r}, \text{ for } j,l = 1,2,3 \quad (59\text{-}1)$$

$$\frac{V^2}{c^2} = -2\frac{M}{r} - (\frac{M}{r})^2 \quad (59\text{-}2)$$

$$c^2 v_l R_l^j + V^j = -\frac{M}{r} \cdot V^j, \text{ for } j,l = 1,2,3 \qquad (59\text{-}3)$$

$$c^2 v^2 = -2\frac{M}{r} - (\frac{M}{r})^2 \qquad (59\text{-}4)$$

Their linear approximation forms are:

$$\frac{V^l \cdot R_j^l}{c^2} + v_j = \frac{M}{r}, \text{ for } j,l = 1,2,3 \qquad (60\text{-}1)$$

$$\frac{V^2}{c^2} = -2\frac{M}{r} \qquad (60\text{-}2)$$

$$c^2 v_l R_l^j + V^j = -\frac{M}{r} \cdot V^j, \text{ for } j,l = 1,2,3 \qquad (60\text{-}3)$$

$$c^2 v^2 = -2\frac{M}{r} \qquad (60\text{-}4)$$

For planets, such as the Earth, the rotation velocity is completely determined by the gravity field ($M$ is negative). For the electrons, the rotation velocity is completely determined by the total charge of electrons. On the other hand, the negative Newtonian matter ($M$ is positive) and the positive Coulomb charge (positive electro or nucleus) have no macro rotation. Please note this conclusion is limited to conservative field.

For the negative Newtonian matter ($M$ is positive) and the positive Coulomb charge (positive electro or nucleus), if they have rotation, their momentum must be imaginary. For such a case, the matter must be in spatial local form, that is:

$$\frac{\partial \tilde{u}^4}{\partial t} = \frac{M}{r} \cdot \exp(\pm \vec{k} \cdot \vec{r}) \qquad (61)$$

In quantum mechanics, such a solution corresponds to the weak interaction of nucleus. Based on this research, the weak interaction has no macro rotation representation. Hence, the weak interaction is not directly observable by kinetic motion. In fact, the weak interaction is inferred from the quantum effects.

For the non-linear exact form (60), what solutions are possible is determined from geometric consideration. The basic conclusion is that: some matter can have rotational motion, some matter cannot have rotational motion. In quantum mechanics, some particles have self-rotation, but some particles have no self-rotation.

Please note that, the rotation is determined by the matter itself, so its intrinsic rotation.

Generally speaking from above research, any Newtonian matter planet has self-rotation. Therefore, the cosmic world is a self-rotation world.

*(B). Quantum Wave Matter*

The quantum wave matter can be introduced by the form solution of time displacement as:

$$\tilde{u}^4 = a \cdot \exp[(\vec{k} \cdot \vec{r} - \omega t)\tilde{j}] \qquad (62)$$

In this research, the physical component is defined by the real parts. As usual custom, this point will not be mentioned in related equations.

Then, when all parameters are constants in four-dimensional spacetime, one gets:

$$\eta = 1 + \frac{\partial \tilde{u}^4}{\partial t} = 1 - \tilde{j}\omega \cdot \tilde{u}^4 \qquad (63)$$

$$v_i = \frac{\partial \tilde{u}^4}{\partial x^i} = \tilde{j} k_i \tilde{u}^4 \tag{64}$$

For the near-flat spacetime, therefore, the geometrical equations can be rewritten as:

$$V^l \cdot R_j^l = -c^2 \cdot \tilde{j} k_j \tilde{u}^4 - c^2 \omega k_j (\tilde{u}^4)^2 , \text{ for } j,l = 1,2,3 \tag{65-1}$$

$$V^2 = c^2 \cdot [2\tilde{j}\omega\tilde{u}^4 + \omega^2 (\tilde{u}^4)^2] \tag{65-2}$$

$$\frac{1}{c^2} \cdot V^j = -\tilde{j}(k_l R_l^j - \omega)\tilde{u}^4 , \text{ for } j,l = 1,2,3 \tag{65-3}$$

$$\omega = \frac{\tilde{j}}{2}(\omega^2 - c^2 k^2)\tilde{u}^4 \tag{65-4}$$

They show that the macro behavior of quantum matter wave is determined by the wave function completely. In quantum mechanics, the frequency and wave-number are defined as:

$$\omega = \frac{E}{\hbar}, \quad \vec{k} = \frac{\vec{p}}{\hbar} \tag{66}$$

Putting them into the equations (65), one gets:

$$V^l \cdot R_j^l = -\frac{c^2}{\hbar} \cdot \tilde{j} p_j \tilde{u}^4 - \frac{c^2}{\hbar^2} E p_j (\tilde{u}^4)^2 , \text{ for } j,l = 1,2,3 \tag{67-1}$$

$$V^2 = c^2 \cdot [\frac{2\tilde{j}}{\hbar} E \tilde{u}^4 + \frac{E^2}{\hbar^2}(\tilde{u}^4)^2] \tag{67-2}$$

$$\frac{1}{c^2} \cdot V^j = -\frac{\tilde{j}}{\hbar}(p_l R_l^j - E)\tilde{u}^4 , \text{ for } j,l = 1,2,3 \tag{67-3}$$

$$E = \frac{\tilde{j}}{2\hbar}(E^2 - c^2 p^2)\tilde{u}^4 \tag{67-4}$$

Their linear approximation form about wave matter is:

$$V^l \cdot R_j^l = -\frac{c^2}{\hbar} \cdot \tilde{j} p_j \tilde{u}^4 , \text{ for } j,l = 1,2,3 \tag{68-1}$$

$$V^2 = c^2 \cdot \frac{2\tilde{j}}{\hbar} E \tilde{u}^4 \tag{68-2}$$

$$\frac{1}{c^2} \cdot V^j = -\frac{\tilde{j}}{\hbar}(p_l R_l^j - E)\tilde{u}^4 , \text{ for } j,l = 1,2,3 \tag{68-3}$$

$$E = \frac{\tilde{j}}{2\hbar}(E^2 - c^2 p^2)\tilde{u}^4 \tag{68-4}$$

For a given wave matter, by these equations the macro motion is completely determined. Based this understanding, the quantum matter motion is deterministic rather than probability-type non-deterministic.

As the energy and wave function are not zero, the above equations show that the macro velocity and macro local rotation are always non zero. Therefore, the macro motion is measurable for quantum matter motion.

In fact, the equations (48) and (49) show that the macro velocity $V^i$ and the micro time displacement spatial gradient $v_i$ are reciprocal items. In this sense, the quantum matter has particle representation and matter wave representation.

**5. Cosmic World as Mass-less Matter Rotation**

As the parameter $\mu$ is related with the Newtonian mass, the mass-less matter is defined as: $\mu = 0$. Here, such a kind of field is named as cosmic field. For the mass-less matter, the related motion equations are:

$$\lambda \frac{\partial R_j^j}{\partial x^i} + \lambda \frac{\partial^2 \tilde{u}^4}{\partial t \partial x^i} = f^i, \quad i, j = 1,2,3 \tag{69-1}$$

$$\frac{\lambda}{c} \frac{\partial R_j^j}{\partial t} + \frac{\lambda}{c} \frac{\partial^2 \tilde{u}^4}{(\partial t)^2} = f^4, \quad j = 1,2,3 \tag{69-2}$$

$$\lambda \frac{\partial R_j^j}{\partial x^i} + \lambda \frac{\partial^2 \tilde{u}^4}{\partial t \partial x^i} = f_i, \quad i, j = 1,2,3 \tag{69-3}$$

$$\frac{\lambda}{c} \frac{\partial R_j^j}{\partial t} + \frac{\lambda}{c} \frac{\partial^2 \tilde{u}^4}{(\partial t)^2} = f_4, \quad j = 1,2,3 \tag{69-4}$$

For cosmic field matter, it is easy to find that:

$$f^i = f_i, \quad i = 1,2,3,4 \tag{70}$$

As by the definition equation (5), one has:

$$R_j^j = 1 + 2\cos\Theta \tag{70}$$

The motion equations become:

$$2\lambda \frac{\partial \cos\Theta}{\partial x^i} + \lambda \frac{\partial^2 \tilde{u}^4}{\partial t \partial x^i} = f^i, \quad i, j = 1,2,3 \tag{71-1}$$

$$\frac{2\lambda}{c} \frac{\partial \cos\Theta}{\partial t} + \frac{\lambda}{c} \frac{\partial^2 \tilde{u}^4}{(\partial t)^2} = f^4 \tag{71-2}$$

Or in the solution form:

$$\frac{\partial \tilde{u}^4}{\partial t} = \frac{Action}{\lambda} - 2\cos\Theta \tag{72}$$

For zero action, one gets:

$$\frac{\partial \tilde{u}^4}{\partial t} = -2\cos\Theta \tag{73}$$

It shows that pure space rotation is equivalent with a homogenous field.

Such a kind of matter may be the so-called dark matter as it is non-visible under the sense of Newtonian matter has mass. Or it may be interpreted as cosmic microwave radiation matter.

Note that when the rotation angular has spatial variation, it is equivalent with a static electrical potential field. When the rotation angular is time dependent, magnetic-like field will be produced.

The related geometric equations are:

$$V^l \cdot R_j^l = -c^2 \cdot (1 - 2\cos\Theta) \cdot v_j, \quad \text{for} \quad j,l = 1,2,3 \tag{74-1}$$

$$V^2 = c^2 \cdot [1 - (1 - 2\cos\Theta)^2] \tag{74-2}$$

$$v_l R_l^j = -\frac{1}{c^2}(1 - 2\cos\Theta) \cdot V^j, \quad \text{for} \quad j,l = 1,2,3 \tag{74-3}$$

$$v^2 = \frac{1}{c^2}[1-(1-2\cos\Theta)^2]  \qquad (74\text{-}4)$$

Geometrically, cosmic field can be described by the curvature spacetime, such as Minkowski world.

Therefore, the cosmic can be described by the equation (74) with one parameter $\Theta$, as an pure rotational space. If one views that the Lorentz rotation group as such a kind of matter, then this matter is identified by the effectiveness of Lorentz transformation.

For such a kind of matter, its motion respect with absolute spacetime (ether) is determined by parameter $\lambda$, this parameter can not be detected by the spacetime curvature. To detect this parameter, strong action is required. On this sense, the parameter $\lambda$ is action related.

In fact, in previous research, when $Action(\widetilde{F}_j^i)$ is taken as the reference matter, the parameters action-dependent of reference matter is implied.

Summering above discussion, the Lorentz rotational invariance group indeed represents the general feature of cosmic spacetime. So, the basis of general relativity is supported by this research. However, this research does not support such a world as the unique solution of cosmic world.

When the matter deformations are highly shear, the $R_j^i$ should be replaced by $\frac{1}{\cos\theta}\widetilde{R}_j^i$. The related equations can be modified accordingly. Generally, for many body problems, this orthogonal definition is essential.

## 6. Cosmic-Quantum Field

For general matter, when the initial matter (taken as the reference matter) is isotropic, the motion equation can be written as:

$$\lambda \frac{\partial F_j^j}{\partial x^i} + \lambda \frac{\partial^2 \widetilde{u}^4}{\partial t \partial x^i} + \mu \frac{\partial F_j^i}{\partial x^j} + \frac{\mu}{c^2}\frac{\partial V^i}{\partial t} = f^i, \quad i,j=1,2,3 \qquad (75\text{-}1)$$

$$\frac{\lambda}{c}\frac{\partial F_j^j}{\partial t} + \mu c \frac{\partial^2 \widetilde{u}^4}{\partial x^j \partial x^j} + \frac{\lambda+\mu}{c}\frac{\partial^2 \widetilde{u}^4}{(\partial t)^2} = f^4, \quad j=1,2,3 \qquad (75\text{-}2)$$

And:

$$\lambda \frac{\partial F_j^j}{\partial x^i} + \lambda \frac{\partial^2 \widetilde{u}^4}{\partial t \partial x^i} + \mu \frac{\partial F_i^j}{\partial x^j} + \mu \frac{\partial^2 \widetilde{u}^4}{\partial x^i \partial t} = f_i, \quad i,j=1,2,3 \qquad (76\text{-}1)$$

$$\frac{\lambda}{c}\frac{\partial F_j^j}{\partial t} + \frac{\mu}{c}\frac{\partial V^j}{\partial x^j} + \frac{\lambda+\mu}{c}\frac{\partial^2 \widetilde{u}^4}{(\partial t)^2} = f_4, \quad j=1,2,3 \qquad (76\text{-}2)$$

When the vacuum is taken as the reference matter, there are four typical cases.

*(A). Pure Newtonian Matter ($\lambda=0$)*

The motion equations are:

$$\mu \frac{\partial F_j^i}{\partial x^j} + \frac{\mu}{c^2}\frac{\partial V^i}{\partial t} = f^i, \quad i,j=1,2,3 \qquad (77\text{-}1)$$

$$\mu c \frac{\partial^2 \widetilde{u}^4}{\partial x^j \partial x^j} + \frac{\mu}{c}\frac{\partial^2 \widetilde{u}^4}{(\partial t)^2} = f^4, \quad j=1,2,3 \qquad (77\text{-}2)$$

$$\mu \frac{\partial F_i^j}{\partial x^j} + \mu \frac{\partial^2 \tilde{u}^4}{\partial x^i \partial t} = f_i, \quad i,j = 1,2,3 \tag{77-3}$$

$$\frac{\mu}{c} \frac{\partial V^j}{\partial x^j} + \frac{\mu}{c} \frac{\partial^2 \tilde{u}^4}{(\partial t)^2} = f_4, \quad j = 1,2,3 \tag{77-4}$$

For the equation (77-1), when the intrinsic spatial deformation is zero, it is Newton acceleration equation. When the spatial forces are zero, the spatial deformation and acceleration is related by equation:

$$\frac{\partial F_j^i}{\partial x^j} = -\frac{1}{c^2} \frac{\partial V^i}{\partial t}, \quad i,j = 1,2,3 \tag{78}$$

It means that in zero gravity condition, all Newtonian matter has mass-independent motion.

The equation (77-2) in fact is the quantum wave equation for Newtonian matter. When the mass is large enough, it can be simplified as:

$$\mu c \frac{\partial^2 \tilde{u}^4}{\partial x^j \partial x^j} = f^4, \quad j = 1,2,3 \tag{79}$$

It gives out the classical gravity field solution.

The equation (77-3) shows that the spatial deformation is related with an electromagnetic field. This explains the planet electromagnetic field. The equation (77-4) shows that the volume variation will cause the harmonic vibration of time displacement.

*(B). Electric Charge-less Matter Motion ( $\lambda + \mu = 0$ )*

The motion equations are:

$$-\mu \frac{\partial F_j^j}{\partial x^i} - \mu \frac{\partial^2 \tilde{u}^4}{\partial t \partial x^i} + \mu \frac{\partial F_j^i}{\partial x^j} + \frac{\mu}{c^2} \frac{\partial V^i}{\partial t} = f^i, \quad i,j = 1,2,3 \tag{80-1}$$

$$-\frac{\mu}{c} \frac{\partial F_j^j}{\partial t} + \mu c \frac{\partial^2 \tilde{u}^4}{\partial x^j \partial x^j} = f^4, \quad j = 1,2,3 \tag{80-2}$$

$$-\mu \frac{\partial F_j^j}{\partial x^i} - \mu \frac{\partial^2 \tilde{u}^4}{\partial t \partial x^i} + \mu \frac{\partial F_i^j}{\partial x^j} + \mu \frac{\partial^2 \tilde{u}^4}{\partial x^i \partial t} = f_i, \quad i,j = 1,2,3 \tag{80-3}$$

$$-\frac{\mu}{c} \frac{\partial F_j^j}{\partial t} + \frac{\mu}{c} \frac{\partial V^j}{\partial x^j} = f_4, \quad j = 1,2,3 \tag{80-4}$$

When the cosmic force is zero, $f^4 = f_4 = 0$, by equations (80-4) and (80-2), one has:

$$\Delta \approx \Delta_0 \exp(t), \quad \Delta_0 \approx c^2 \frac{\partial^2 \tilde{u}^4}{\partial x^j \partial x^j} \tag{81}$$

Where, $\Delta$ represents the spatial volume of matter, $\Delta_0$ is the spatial volume of matter at zero time.

It shows that the volume of matter will be expanding infinitely. This may be the intrinsic explanation about the Big-ban when the matter is in cosmic background (scale).

*(C). Strong Electric Charge Matter Motion ( $\lambda = \mu$ )*

The motion equations are:

$$\mu \frac{\partial F_j^j}{\partial x^i} + \mu \frac{\partial^2 \tilde{u}^4}{\partial t \partial x^i} + \mu \frac{\partial F_j^i}{\partial x^j} + \frac{\mu}{c^2} \frac{\partial V^i}{\partial t} = f^i, \quad i,j = 1,2,3 \tag{82-1}$$

$$\frac{\mu}{c} \frac{\partial F_j^j}{\partial t} + \mu c \frac{\partial^2 \tilde{u}^4}{\partial x^j \partial x^j} + \frac{2\mu}{c} \frac{\partial^2 \tilde{u}^4}{(\partial t)^2} = f^4, \quad j = 1,2,3 \tag{82-2}$$

$$\mu \frac{\partial F_j^{\,j}}{\partial x^i} + \mu \frac{\partial^2 \widetilde{u}^{\,4}}{\partial t \partial x^i} + \mu \frac{\partial F_i^{\,j}}{\partial x^j} + \mu \frac{\partial^2 \widetilde{u}^{\,4}}{\partial x^i \partial t} = f_i, \quad i, j = 1,2,3 \tag{82-3}$$

$$\frac{\mu}{c} \frac{\partial R_j^{\,j}}{\partial t} + \frac{\mu}{c} \frac{\partial V^{\,j}}{\partial x^j} + \frac{2\mu}{c} \frac{\partial^2 \widetilde{u}^{\,4}}{(\partial t)^2} = f_4, \quad j = 1,2,3 \tag{82-4}$$

When the cosmic force is zero, $f^4 = f_4 = 0$, and $\frac{2\mu}{c} \frac{\partial^2 \widetilde{u}^{\,4}}{(\partial t)^2} = 0$, by equations (82-4) and (82-2), one has:

$$\Delta \approx \Delta_0 \exp(-t), \quad \Delta_0 \approx c^2 \frac{\partial^2 \widetilde{u}^{\,4}}{\partial x^j \partial x^j} \tag{83}$$

Where, $\Delta$ represents the spatial volume of matter, $\Delta_0$ is the spatial volume of matter at zero time.

It shows that the volume of matter will be contracting infinitely. This may be the intrinsic explanation about the Black-hole when the matter is in cosmic background (scale).

## 6. Conclusion

The rotational motion is discussed in this part. The related geometrical equations form the intrinsic geometrical equations for quantum mechanics in finite geometrical field theory of matter motion. For the Newtonian matter rotation in vacuum or ether, it can be concluded that the Euler rotation equation for rigid body is the only rotation mode when the reference matter is vacuum or Cartesian spacetime. Therefore, for curvature spacetime, that is, when the real cosmic field is in consideration, the related equations (can be used to explain the so-called self rotation of spaceship, and so can be used to control to flight of spaceship) may be valuable for astronomy research and cosmic study. The shear-wave-like feature of electromagnetic field is explained.

In this research, the reference matter invariance, that is the invariant of $\eta$, is equivalent with the proper time invariant in general relativity. In fact, this equation explains the true physical meaning of Lorentz transformation. Please remember that the Lorentz electromagnetic wave equation is taken the cosmic background as the reference matter. Generally speaking from the research, any Newtonian matter planet has self-rotation. Therefore, the cosmic world is a self-rotation world. Weak interaction is introduced as an approximation.

The general matter motion is discussed, the main conclusions are:

(1). For the case $\mu = 0$, there exists a space rotation field. For this matter field, the Lorentz rotational invariance group indeed represents the general feature of cosmic spacetime. So, the basis of general relativity is supported by this research. However, this research does not support such a world as the unique solution of cosmic world. The research shows that such a kind of matter may be the so-called dark matter as it is non-visible under the sense of Newtonian matter has mass. Or it may be interpreted as cosmic microwave radiation matter. The research discovers that when the rotation angular has spatial variation, it is equivalent with a static electrical potential field. When the rotation angular is time dependent, magnetic-like field will be produced. Geometrically, cosmic vacuum field can be described by the curvature spacetime, such as Minkowski world.

(2). For the case $\lambda = 0$, it gives out the classical gravity field solution. The research also shows that the spatial deformation is related with an electromagnetic field. This explains the planet electromagnetic field. Further more, the research shows that the volume variation will cause the harmonic vibration of time displacement.

(3). For the case $\lambda = -\mu$, it shows that the volume of matter will be expanding infinitely. This may be the intrinsic explanation about the Big-ban when the matter is in cosmic background (scale).

(4). For the case $\lambda = \mu$, it shows that the volume of matter will be contracting infinitely. This may be the intrinsic explanation about the Black-hole when the matter is in cosmic background (scale).

**Reference**

[1] Xiao Jianhua. *Inertial System and Special Relativity-Finite Geometrical Field Theory of Matter Motion Part One.* arXiv: physics/0512110, 2005, 1-19

[2] Xiao Jianhua. *Gravity Field and Electromagnetic Field-Finite Geometrical Field Theory of Matter Motion Part Two.* arXiv: physics/0512123, 2005, 1-16

[3]. Chen Zhida. *Rational Mechanics—Non-linear Mechanics of Continuum.* Xizhou: China University of Mining & Technology Publication, (In Chinese) 1987

[4]. Chen Zhida. *Rational Mechanics.* Chongqin: Chongqin Publication, (In Chinese) 2000

[5] Xiao Jianhua, Decomposition of displacement gradient and strain definition, *Advance in Rheology and its Application(2005),* Science Press USA Inc, 2005, 864-868

[6] Einstein, A., Relativistic Theory of Non-Symmetric Field, In *Einstein's Work Collection (Vol.2),* Commercial Pub., 1977, p560-565 (in Chinese). (Original: Einstein, A., Meaning of Relativity, (5[th] edition), 1954, p133-166).

[7] Dittrich, W., M. Reuter, *Classical and Quantum Dynamics (Second ed.),* New York: Springer-Verlag, 1996

[8] Dubrovin, B. A., A. T. Femenko, S. P. Novikov, *Modern Geometry—Methods and Application, Part I: The Geometry of Surfaces, Transformation Groups and Fields.* New York: Springer-Verlag, 1984

[9] Zeng Jingyuan. Quantum Mechanics (Vol 1 & 2), third ed. Beijing: Science Publication, 2000

**Appendix A    Mathematic Feature of Deformation Tensor**

For continuum, each material point can be parameterized with continuous coordinators $x^i, i = 1,2,3$. When the coordinators are fixed for each material point no matter what motion or deformation happens, the covariant gauge field $g_{ij}$ at time $t$ will define the configuration in the time. The continuous coordinators endowed with the gauge field tensor define a co-moving dragging coordinator system.

The initial configuration gauge $g_{ij}^0$ defines a distance geometric invariant:

$$ds_0^2 = g_{ij}^0 dx^i dx^j \tag{A-1}$$

The symmetry and positive feature of gauge tensor insures that there exist three initial covariant base vectors $\vec{g}_i^0$ make:

$$g_{ij}^0 = \vec{g}_i^0 \cdot \vec{g}_j^0 \tag{A-2}$$

For current configuration, three current covariant base vectors $\vec{g}_i$ exist which make:

$$g_{ij} = \vec{g}_i \cdot \vec{g}_j \tag{A-3}$$

The current distance geometric invariant is:

$$ds^2 = g_{ij} dx^i dx^j \tag{A-4}$$

For each material point, there exists a local transformation $F_j^i$, which relates the current covariant base vectors with initial covariant base vectors:

$$\vec{g}_i = F_i^j \vec{g}_j^0 \tag{A-5}$$

So, the current covariant gauge tensor can be expressed as:

$$g_{ij} = F_i^k F_j^l g_{kl}^0 \tag{A-6}$$

In Riemann geometry, contra-variant gauge tensor $g^{ij}$, $g^{0ij}$ can be introduced, which meets condition:

$$g^{il} g_{jl} = \delta_j^i, \quad g^{0il} g_{jl}^0 = \delta_j^i \tag{A-7}$$

Similarly, contra-variant base vectors $\vec{g}^i$, $\vec{g}^{0i}$ can be introduced for current configuration and initial configuration respectively. Mathematically, there are:

$$g^{ij} = \vec{g}^i \cdot \vec{g}^j, \quad g^{0ij} = \vec{g}^{0i} \cdot \vec{g}^{0j} \tag{A-8}$$

There exists a local transformation $G_j^i$, which relates the contra-variant base vectors:

$$\vec{g}^i = G_j^i \vec{g}^{0j} \tag{A-9}$$

So, the current contra-variant gauge tensor can be expressed as:

$$g^{ij} = G_k^i G_l^j g^{0kl} \tag{A-10}$$

By equations (A-6), (A-7), and (A-10), it is easy to find out that:

$$G_l^i F_j^l = \delta_j^i \tag{A-11}$$

Hence, the transformation $F_j^i$ relates the initial contra-variant base vectors with current contra-variant base vectors in such a way that:

$$\vec{g}^{0i} = F_j^i \vec{g}^j \tag{A-12}$$

Therefore, the transformation $F_j^i$ is a mixture tensor. Its lower index represents covariant component in $g_{ij}^0$ configuration, its upper index represents contra-variant component in $g_{ij}^0$ configuration.

Similar discussion shows that the local transformation $G_j^i$ is a mixture tensor, lower index represents covariant component in $g_{ij}$ configuration, upper index represents contra-variant component in $g_{ij}$ configuration. It is easy to find that:

$$\vec{g}_i^0 = G_i^j \vec{g}_j \tag{A-13}$$

Other two important equations are:

$$F_j^i = \vec{g}^{0i} \cdot \vec{g}_j \tag{A-14}$$

$$G_j^i = \vec{g}^i \cdot \vec{g}_j^0 \tag{A-15}$$

Bt these equations, $F_j^i$ can be explained as the extended-Kronecker-delta in that it's the dot product of contra-variant base vector in initial configuration and covariant base vector in current configuration. When the two configurations are the same, it becomes the standard Kronecker-delta. For the $G_j^i$, similar interpretation can be made.

From these equations to see, geometrically, it is found that the $F_j^i$ measures the current covariant base vector with the initial contra-variant base vector as reference. When stress tensor is defined as:

$$\sigma_j^i = E_{jl}^{ik}(F_k^l - \delta_k^l) \tag{A-16}$$

Its mechanic meaning can be explained as the lower index represents component of surface force in current direction, and the upper index represents the initial surface normal which surface force acts on. It takes the initial surface as surface force reference.

The $G_j^i$ measures the initial covariant base vector with the current contra-variant base vector as reference. The corresponding stress can be explained as the lower index represents component of surface force in initial direction, and the upper index represents the current surface normal which surface force acts on.

For such a kind of mixture tensor, it is defined on the same point with different configurations (that is initial and current). So, the name of two-point tensor given by C Truessdell is not correct. This mis-interpretation has caused many doubts cast on the feature of transformation tensor. Some mathematician even said that the mixture tensor is meaningless.

However, during treating the non-symmetric field theory, Einstein believes that the use of mixture tensor is more reasonable that the pure covariant or pure contra-variant tensor. Recently, the concept of bi-tensor is introduced in physics topic. So, we have sound reason to use mixture tensor in continuum mechanics, as it can give clear physical meaning for the definition of stress.

Note that the transformation $F_j^i$ is completely determined by the deformation measured in initial configuration with gauge tensor $g_{ij}^0$. Mathematically, the covariant differentiation is taken in the initial geometry also, although the physical meaning of $F^i_j$ is that it relates two configurations.

It is valuable to point out that, if the initial coordinator system is taken as Cartesian system, the $F^i_j$ can be transformed into pure covariant form:

$$F_{ij} = \vec{g}^0_i \cdot \vec{g}_j \qquad (A\text{-}17)$$

The $G^i_j$ can be transformed into pure contra-variant form:

$$G^{ij} = \vec{g}^i \cdot \vec{g}^{0j} \qquad (A\text{-}18)$$

Although it is acceptable in form for the special case of taking Cartesian system as the initial coordinator system, the intrinsic meaning of deformation tensor is completely destroyed by such a formulation. That may be the main reason for the rational mechanics constructed by C Truessdell et al in 1960s.

Mathematically, once the initial gauge field is selected, the current gauge field is to be obtained by the given physical deformation. In this sense, the current gauge field is viewed as the physical field. So, the covariant differentiation is taken respect with the initial configuration.

Truessdell argued that the covariant differentiation should be taken one index in initial configuration, another index in current configuration. This concept had strongly effects on the development of infinite deformation mechanics. Such a kind of misunderstanding indeed is caused by the equations (A17-18).

Historically, Chen Zhida is the first one to clear the ambiguity caused by equations (A-17) and (A-18) systematically. His monograph "Rational Mechanics"(1987) treats this topic in depth.

There are many critics about the mathematics used in Chen's rational mechanics theory. The most comment one is that: for a coordinator transformation

$$dx^i = A^i_j dX^j, \quad dX^i = (A^i_j)^{-1} dx^j \qquad (A\text{-}19)$$

the covariant and contra-variant components of tensor is defined by its transformation from new coordinator system to old coordinator system follows $A^i_j$ or $(A^i_j)^{-1}$ formulations. However, such a tensor definition is to make:

$$ds^2 = g_{ij} dx^i dx^j = G_{ij} dX^i dX^j \qquad (A\text{-}20)$$

be invariant. Such a tensor describes a continuum without any motion. In fact, such a tensor feature is only to say the object indifference for coordinator system selection.

However, many physicists and mechanists treat motion of continuum as the transformation $A^i_j$ or $(A^i_j)^{-1}$. Mathematically, the equation (A-20) can be rewritten as:

$$ds^2 = g_{ij} dx^i dx^j = g_{kl} A^k_i A^l_j dX^i dX^j = G_{ij} dX^i dX^j \qquad (A\text{-}21)$$

or:

$$ds^2 = G_{ij} dX^i dX^j = G_{kl} (A^k_i)^{-1} (A^l_j)^{-1} dx^i dx^j = g_{ij} dx^i dx^j \qquad (A\text{-}22)$$

It is clear that the covariant invariant feature must be maintained by the coordinator system choice. It has no any meaning of motion. In Chen's geometry, for time parameter $t$,

$$ds^2(t) = g_{ij}(t)dx^i dx^j = g_{kl}(0)F_i^k(t)F_j^l(t)dx^i dx^j \qquad \text{(A-23)}$$

The gauge field is time dependent, while the coordinator is fixed (called intrinsic coordinator).

The equation (A-23) can be rewritten as:

$$ds^2(t) = g_{ij}(t)dx^i dx^j = g_{kl}(0)F_i^k(t)F_j^l(t)dx^i dx^j = g_{kl}(0)dX^k(t)dX^l(t) \qquad \text{(A-24)}$$

It is similar in form with equation (A-21). But their mechanical interpretation is strikingly different.